\documentclass[
prd
,showpacs ,amssymb,nobibnotes,aps,eqsecnum]{revtex4}
\usepackage{graphicx}
\input{epsf}

\usepackage{amsmath,amssymb}
\usepackage{bm}

\newcommand{\dalm}{\kern1pt\vbox{\hrule height 0.9pt\hbox{\vrule width 0.9pt
\hskip 2.5pt\vbox{\vskip 5.5pt}\hskip 3pt\vrule width 0.3pt}\hrule height 0.3pt}
\kern1pt}

\begin{document}

%\twocolumn[\hsize\textwidth\columnwidth\hsize\csname @twocolumnfals\endcsname

% For two column
%\wideabs{

\title{Toroidal Oscillations of slowly rotating relativistic star in tensor-vector-scalar theory }

\author{Hajime Sotani} \email{hajime.sotani@nao.ac.jp}
%\author{Nobutoshi Yasutake$^1$} 
%\author{Toshiki Maruyama$^2$}
%\author{Toshitaka Tatsumi$^3$}
\affiliation{
Division of Theoretical Astronomy, National Astronomical Observatory of Japan, 
2-21-1 Osawa, Mitaka, Tokyo 181-8588, Japan
%$^2$Advanced Science Research Center, Japan Atomic Energy Agency, Tokai, Ibaraki 319-1195, Japan\\
%$^3$Department of Physics, Kyoto University, Kyoto 606-8502, Japan
}

\date{\today}

% Abstract
\begin{abstract}
 We examine the toroidal oscillations on the slowly rotating relativistic stars in tensor-vector-scalar (TeVeS) theory with the Cowling approximation. As a result, we find that perturbation equations describing the toroidal oscillations are same equation form as in general relativity (GR). Although the frequencies of toroidal oscillations in TeVeS are not so different from those in GR, the momentum inertia depends strongly on the gravitational theory. Thus, observing the frequencies of toroidal oscillations and momentum inertia with high accuracy might reveal the gravitational theory in the strong-field regime.
\end{abstract}

%
%%%%%%%%%%%%%%%%%%%%%%%%%%%%%%%%%%%%%%%%%%%%%%%%%
%  04.40.Dg :  Relativistic stars: structure, stability, and oscillations (see also 97.60.-s Late stages of stellar evolution) 
%  04.50.Kd : Modified theories of gravity
%  04.80.Cc : Experimental tests of gravitational theories 
%  97.60.Jd :  Neutron stars (see also 26.60.+c Nuclear matter aspects of neutron stars in nuclear physics) 
%%%%%%%%%%%%%%%%%%%%%%%%%%%%%%%%%%%%%%%%%%%%%%%%%
%]
% For two column
%}
\maketitle
%\baselineskip 24pt
%%%%%%%%%%%%%%%%%%%%%%%%%%%%%%%%%%%%%%%%%%%%%%%%
\section{Introduction}
\label{sec:I}
%%%%%%%%%%%%%%%%%%%%%%%%%%%%%%%%%%%%%%%%%%%%%%%%

Via many experiments, the validity of general relativity (GR) has been shown in the weak gravitational field such as solar system, while the tests of gravitational theory in the strong-field regime are still very poor. Accordingly, the gravitational theory in the strong gravitational field has not been constrained observationally. However, with the development of technology, it is becoming possible to observe the compact objects with high accuracy and via these observations it could be possible to test the gravitational theory in the strong gravitational field \cite{Psaltis2009}. In practice, some possibilities to distinguish the gravitational theory are suggested. For example, using the surface atomic line redshifts \cite{DeDeo2003} or gravitational waves radiated from the neutron stars \cite{Sotani2004}, it could be possible to distinguish the scalar-tensor theory proposed in \cite{Damour1992} from GR. Additionally, Corda suggested the definitive test for GR with gravitational waves\cite{Corda2009}.

The tensor-vector-scalar (TeVeS) theory, which is originally proposed by Bekenstein \cite{Bekenstein2004}, is one of the alternative gravitational theory and attracts considerable attention. This theory is covariant formalism for modified Newtonian dynamics \cite{Milgrom1983,Skordis2009}. The reason why this theory gets attention is possible to explain the galaxy rotational curve and Tully-Fisher law without the presence of dark matter \cite{Bekenstein2004}. TeVeS is also successfully to explain the strong gravitational lensing \cite{Chen2006} as well as the galaxy distributions through an evolving Universe without cold dark matter \cite{Dodelson2006}. It should be noticed that the bullet cluster 1E0657-558 might be good candidate to make a constraint in the gravitational theory observationally \cite{Brownstein2007}, but so far no one has tried this attempt for TeVeS yet. Drawing attention to the strong gravitational region of TeVeS, Giannios found the Schwarzschild solution \cite{Giannios2005}, Sagi and Bekenstein found the Reissner-Nordstr\"om solution \cite{Sagi2008}, and Lasky $et$ $al.$ derived the Tolman-Oppenheimer-Volkoff (TOV) equations in TeVeS and produced the static, spherically symmetric stellar models in TeVeS \cite{Paul2008}. Recently, Lasky and Donova examined the stability and quasi-normal modes of black hole in TeVeS \cite{Paul2010}.

Additionally, it has been suggested how to distinguish TeVeS from GR observationally. For example, one could reveal the gravitational theory in the strong-field regime with the redshift of the atomic spectral lines emanating from the surface of neutron star \cite{Paul2008}, with the Shapiro delays of gravitational waves and photons or neutrinos \cite{Desai2008}, and with the spectrum of gravitational waves emitted from the compact objects \cite{Sotani2009}, where the compact objects were assumed to be spherically symmetry. On the other hand, in this article, we will focus on the toroidal oscillations in the slowly rotating compact objects constructed in \cite{Sotani2010}, which are associated with the $r$ modes gravitational waves. The $r$ modes arise due to the rotational effects and degenerate into zero frequency in the limit of non-rotatioin.

The observations of stellar oscillations via gravitational waves are considered to provide a unique tool to estimate the stellar parameters such as mass, radius, rotation rate, magnetic fields, and equation of state (e.g., \cite{AK1996,Sotani2001,Sotani2003,Miltos2007,Erich2009}), which is called ``gravitational wave asteroseismology". The detailed analysis of the gravitational waves also makes it possible to determine the radius of accretion disk around supermassive black hole \cite{Sotani2006} or to know the magnetic effect during the stellar collapse \cite{Sotani2007a}.

 In this article, as a first step to see the dependence of toroidal oscillations on the gravitational theory, we assume the Cowling approximation, i.e., we see only fluid oscillations and the perturbations of the other fields will be omitted. The more detailed study including the oscillations of the other fields will be done near future. This article is organized as follows. In the next section, we review the fundamental parts of TeVeS and the stellar model with slow rotation in TeVeS. In Sec. \ref{sec:III}, we derive the perturbation equations describing the toroidal oscillations with the Cowling approximation and present the frequencies of toroidal oscillations as varying the stellar parameters. Finally, we make a conclusion in Sec. \ref{sec:IV}. In this article, we adopt the unit of $c=G=1$, where $c$ and $G$ denote the speed of light and the gravitational constant, respectively, and the metric signature is $(-,+,+,+)$.

%%%%%%%%%%%%%%%%%%%%%%%%%%%%%%%%%%%%%%%%%%%%%%%%
\section{Stellar Models in TeVeS}
\label{sec:II}
%%%%%%%%%%%%%%%%%%%%%%%%%%%%%%%%%%%%%%%%%%%%%%%%
%%%%%%%%%%%%%%%%%%%%%%%%%%%%%%%%%%%%%%%%%%%%%%%%
\subsection{TeVeS}
\label{sec:II-1}
%%%%%%%%%%%%%%%%%%%%%%%%%%%%%%%%%%%%%%%%%%%%%%%%

The details of TeVeS can be found in \cite{Bekenstein2004}. TeVeS is based on three dynamical gravitational fields; an Einstein metric $g_{\mu\nu}$, a timelike four-vector field ${\cal U}^\mu$, and a scalar field $\varphi$. Additionally there is a nondynamical scalar field $\sigma$. The vector field satisfy the normalization condition with Einstein metric as $g_{\mu\nu}{\cal U}^\mu{\cal U}^\nu=-1$ and the physical metric $\tilde{g}_{\mu\nu}$ is defined with the Einstein metric as
\begin{equation}
 \tilde{g}_{\mu\nu} = e^{-2\varphi}g_{\mu\nu} - 2{\cal U}_\mu{\cal U}_\nu\sinh(2\varphi).
\end{equation}
The total action of TeVeS, $S$, contains contributions from the three dynamical fields mentioned the above and a matter contribution (see \cite{Bekenstein2004} for the details). This total action has two positive dimensionless parameters $k$ and $K$, which are corresponding to the coupling parameters for the scalar and vector fields, respectively. Varying the total action $S$ with respect to $g^{\mu\nu}$, ${\cal U}_\mu$, and $\varphi$, one can get the field equations for the tensor, vector, and scalar fields (see \cite{Bekenstein2004} for the explicit field equations). Since the previous study about the neutron star structure in TeVeS has shown that the stellar properties are almost independent from the scalar coupling $k$ \cite{Paul2008}, in this article we focus only on the dependence of vector coupling $K$. Although, the restrictions on $K$ have not been discussed in great detail in the literature, in \cite{Paul2008} they showed that to construct the stellar models $K$ has to be less than 2 and also that $K$ should be less than 1 to produce a realistic stellar mass. Thus, in this article we examine as varying $K$ in the range of $0<K\le 1$.

%%%%%%%%%%%%%%%%%%%%%%%%%%%%%%%%%%%%%%%%%%%%%%%%
\subsection{Slowly Rotating Relativistic Stellar Models}
\label{sec:II-2}
%%%%%%%%%%%%%%%%%%%%%%%%%%%%%%%%%%%%%%%%%%%%%%%%

As a background stellar model, we consider a slowly rotating relativistic star with a uniform angular velocity $\tilde{\Omega}$, where the rotational axis is set to be $\theta=0$. Since the details for constructing such stellar models have shown in \cite{Sotani2010}, in this section we describe only essential points. In the framework for slow rotation, we assume to keep only the linear effects in the angular velocity. Then, the stellar models are still spherical, because the deformation due to the rotation is of the order $\tilde{\Omega}^2$. Those stellar models in TeVeS can be constructed by using the recipe shown in \cite{Paul2008} and the metric in physical frame is given by
\begin{equation}
 d\tilde{s}^2 = -e^{\nu+2\varphi} dt^2 + e^{\zeta-2\varphi}dr^2 + r^2e^{-2\varphi}\left(d\theta^2 + \sin^2\theta d\phi^2\right) 
                         - 2\omega r^2 e^{-2\varphi} \sin^2\theta dtd\phi,
\end{equation}
where $\nu$, $\zeta$, and $\omega$ are functions of the radial coordinate $r$. Up to first order of $\tilde{\Omega}$, the background fluid four-velocity of the star is described as
\begin{equation}
 \tilde{u}^\mu = \left[e^{-\varphi-\nu/2},\,0,\,0,\,\tilde{\Omega}e^{-\varphi-\nu/2}\right].
\end{equation}
In TeVeS, another variable also needs to determine the rotational dragging, i.e., induced vector field due to the rotation, which is described as $\delta{\cal U}^\phi=-{\cal V}(r)$ \cite{Sotani2010}. With appropriate boundary conditions at the stellar center and infinity, by calculating two second order differential equations with respect to $\omega$ and ${\cal V}$ as shown in \cite{Sotani2010}, one can determine the distribution of rotational frame dragging. About the stellar matter, we assume the perfect fluid described by the energy-momentum tensor
\begin{equation}
 \tilde{T}^{\mu\nu} = \left(\tilde{\rho} + \tilde{P}\right)\tilde{u}^\mu\tilde{u}^\nu + \tilde{P}\tilde{g}^{\mu\nu},
\end{equation}
where $\tilde{\rho}$ and $\tilde{P}$ are the energy density and pressure in the physical frame, respectively. The adopted equations of state to construct the stellar model are the same ones as in \cite{Sotani2004}, which are polytropic ones derived by fitting functions to tabulated data of realistic equations of state known as EOS A and EOS II.  The maximum masses of neutron stars with these equations of state in GR are $M=1.65M_\odot$ for EOS A  and $M=1.95M_\odot$ for EOS II. That is, EOS A and EOS II are considered as soft and intermediate equations of state, respectively.

%%%%%%%%%%%%%%%%%%%%%%%%%%%%%%%%%%%%%%%%%%%%%%%%%  FIGURE
%%%%%%%%%%%%%%%%%%%%%%%%%%%%%%%%%%%%%%%%%%%%%%%%% Figure 1
\begin{figure}[htbp]
\begin{center}
\includegraphics[scale=0.45]{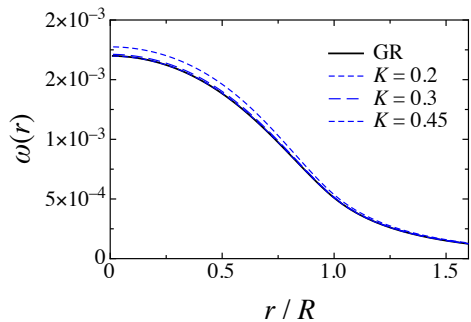} 
\end{center}
\caption{%%
Distribution of $\omega(r)$ in physical frame with different values of $K$ for the stellar model with EOS A and $\tilde{\Omega}=1$ kHz, where stellar mass is fixed to be $M_{\rm ADM}=1.4 M_\odot$. In order to compare the results in TeVeS, the distribution in GR is also plotted with a solid line.
}%%
\label{fig:omega-r}
\end{figure}

As an example, the distribution of $\omega(r)$ in physical frame for the stellar model with $M_{\rm ADM}=1.4M_\odot$ and $\tilde{\Omega}=1$ kHz is plotted in Fig. \ref{fig:omega-r}, where $M_{\rm ADM}$ denotes the total Arnowitt-Deser-Misner (ADM) mass (see \cite{Paul2008} for the definition of ADM mass). In the previous article, it was shown that the distribution of $\omega(r)$ depends on the value of vector coupling, $K$, as well as the total angular momentum of the star $\tilde{J}$, which is determined from the asymptotic behavior of $\omega(r)$ as $\omega(r)=2\tilde{J}/r^2+{\cal O}(1/r^4)$ \cite{Sotani2010}. However, in this article we find that, if one would see the distribution of $\omega$ as a function of $r/R$ where $R$ is stellar radius, those distributions with different value of $K$ are very similar to that in GR.

%%%%%%%%%%%%%%%%%%%%%%%%%%%%%%%%%%%%%%%%%%%%%%%%
\section{Toroidal Oscillations}
\label{sec:III}
%%%%%%%%%%%%%%%%%%%%%%%%%%%%%%%%%%%%%%%%%%%%%%%%

In this article we focus on the toroidal oscillations with the Cowling approximation. That is, we consider only the fluid perturbation with axial parity and the other perturbations of scalar, vector, and tensor fields are neglected. We should notice that the Cowling approximation in GR is typically quite good for toroidal oscillations (axial parity) because this type of oscillations dose not involve the variation of density, while for spheroidal oscillations (polar parity) one can qualitatively discuss those frequencies but the error for typical relativistic stellar models could become less than $20\%$ for the fundamental modes and around $10\%$ for the pressure modes \cite{Yoshida1997}. Similarly GR, we might expect the validity of the Cowling approximation for toroidal oscillations in TeVeS. However, as pointed out in \cite{Sotani2010}, for the slowly rotating star in TeVeS there exists the induced vector field due to the rotation and this induced vector field might play an important role in the toroidal oscillations, when one considers the metric perturbation. As a future work, we will examine the toroidal oscillations with the perturbations of the other fields.

With the approximation of slowly rotating star as mentioned in Sec. \ref{sec:II}, we consider the rotational effect up to first order of $\tilde{\Omega}$. On this slowly rotating star, we add the fluid oscillations. Then, the leading order in the linearized field equations becomes $\sim{\cal O}(\tilde{\Omega}\epsilon)$, where $\epsilon$  expresses the order of fluid oscillations. Additionally, one can see that the coupling between the toroidal and spheroidal oscillation becomes higher order effects than ${\cal O}(\tilde{\Omega}\epsilon)$  (see \cite{Kojima1998} for the ordering in GR). In this article, as a first step, we take into account the leading order of the fluid oscillations, in which we omit the coupling between the toroidal and spheroidal oscillations. Considering the toroidal oscillations, the Lagrangian displacement vector for the fluid perturbation is
\begin{equation}
 \tilde{\xi}^i=\left(\tilde{\xi}^r,\tilde{\xi}^\theta,\tilde{\xi}^\phi\right)
      = \left(0,-Z\frac{1}{\sin\theta}\partial_\phi Y_{\ell m},Z\frac{1}{\sin\theta}\partial_\theta Y_{\ell m}\right),
\end{equation}
where $Z$ is a function of $t$ and $r$, while $Y_{\ell m}=Y_{\ell m}(\theta,\phi)$ is the spherical harmonic function. 
Then, the perturbations of four-velocity in the physical frame, $\delta\tilde{u}^\mu$, can be written as
\begin{align}
 \delta \tilde{u}^r &= 0, \\
 \delta \tilde{u}^\theta &= -e^{-\varphi-\nu/2} \partial_t Z \frac{1}{\sin\theta}\partial_\phi Y_{\ell m}, \\
 \delta \tilde{u}^\phi &= e^{-\varphi-\nu/2} \partial_t Z \frac{1}{\sin\theta} \partial_\theta Y_{\ell m}, \\
 \delta \tilde{u}^t &= \left(\tilde{\Omega}-\omega\right)r^2 e^{-4\varphi-\nu}\sin^2\theta\,\delta \tilde{u}^\phi.
\end{align}
It should be noticed that in order to get the expression of $\delta \tilde{u}^t$, we use the relation that $\tilde{g}_{\mu\nu}\tilde{u}^\mu\delta\tilde{u}^\nu=0$, which is obtained the normalization condition for fluid four-velocity, i.e., $\tilde{g}_{\mu\nu}\tilde{u}^\mu\tilde{u}^\nu=-1$.
With the Cowling approximation, the axial perturbation of energy-momentum tensor is given by
\begin{equation}
 \delta\tilde{T}^{\mu\nu} = \left(\tilde{\rho} + \tilde{P}\right)\left(\delta\tilde{u}^\mu\tilde{u}^\nu + \tilde{u}^\mu\delta\tilde{u}^\nu\right).
\end{equation}

The perturbation equation describing the toroidal oscillations can be obtained by taking a variation of the energy-momuntum conservation law, $\tilde{\nabla}_\nu\tilde{T}^{\mu\nu}=0$, which can be reduced as $\tilde{\nabla}_\nu\delta\tilde{T}^{\mu\nu}=0$ with the Cowling approximation. The explicit forms with $\mu=\theta$ and $\phi$ are
\begin{gather}
 \alpha_{\ell m}(t,r)\frac{1}{\sin\theta}\partial_\phi Y_{\ell m} -\beta_{\ell m}(t,r)\cos\theta\partial_\theta Y_{\ell m} =0, \label{eq:theta} \\
 \alpha_{\ell m}(t,r)\partial_\theta Y_{\ell m} + \beta_{\ell m}(t,r)\frac{\cos\theta}{\sin\theta}\partial_\phi Y_{\ell m}=0, \label{eq:phi}
\end{gather}
where the coefficients $\alpha_{\ell m}$ and $\beta_{\ell m}$ are
\begin{align}
 \alpha_{\ell m} &= \partial_t^2 Z + im\tilde{\Omega}\partial_t Z, \\
 \beta_{\ell m} &= 2\left(\omega-\tilde{\Omega}\right)\partial_t Z. 
\end{align}
Calculating $({\rm Eq}. (\ref{eq:theta}))/\sin\theta\,\partial_\phi Y^*_{\ell m}+({\rm Eq}. (\ref{eq:phi}))\partial_\theta Y^*_{\ell m}$ and integrating over the solid angle, one can get the single perturbation equation for the toroidal oscillations, such as
\begin{equation}
 \partial_t^2 Z + im\left[\tilde{\Omega} - \frac{2}{\ell(\ell+1)}\left(\tilde{\Omega}-\omega\right)\right]\partial_t Z=0, \label{eq:perturbation}
\end{equation}
where $Y^*_{\ell m}$ denotes the complex conjugate of $Y_{\ell m}$ and to derive Eq. (\ref{eq:perturbation}) we use the following relations;
\begin{align}
 \int_0^{2\pi}\int_0^{\pi}\left(\partial_\theta Y^*_{\ell m}\partial_\theta Y_{\ell' m'}
      + \frac{1}{\sin^2\theta}\partial_\phi Y^*_{\ell m}\partial_\phi Y_{\ell' m'}\right)\sin\theta d\theta d\phi
      & = \ell\left(\ell+1\right)\delta_{mm'}\delta_{\ell\ell'}, \\
 \int_0^{2\pi}\int_0^{\pi}\left(\frac{\cos\theta}{\sin\theta}\partial_\theta Y^*_{\ell m}\partial_\phi Y_{\ell' m'}
      - \frac{\cos\theta}{\sin\theta}\partial_\phi Y^*_{\ell m}\partial_\theta Y_{\ell' m'}\right)\sin\theta d\theta d\phi 
      & = im\delta_{mm'}\delta_{\ell\ell'},
\end{align}
where $\delta_{\ell\ell'}$ denotes the Kronecker delta. Now, assuming the time dependence of perturbation variable $Z$ as $Z(t,r)=Z(r)e^{i\sigma t}$, the frequencies of toroidal oscillations are given by
\begin{equation}
 \sigma = -m \left[\tilde{\Omega} - \frac{2}{\ell(\ell+1)}\left(\tilde{\Omega}-\omega\right)\right]. \label{eq:sigma0}
\end{equation}
In spite of the fact that the adopted gravitational theory is different from GR, these equations (\ref{eq:perturbation}) and (\ref{eq:sigma0}) are same forms as in GR. That is, at least with the Cowling approximation, the background scalar and vector fields do not affect directly on the toroidal oscillations.

In the Newtonian limit, in which $\omega\to 0$, the spectrum of toroidal oscillations is discrete and the frequency becomes single value as $\sigma=[2/\ell/(\ell+1)-1]m\tilde{\Omega}$. However, when one considers the relativistic effect, i.e., frame dragging, the frequency of toroidal oscillations is not single value. Unlike Newtonian case, the frequencies become the function of $\omega$, i.e., the function of $r$. Namely, similar to GR case \cite{Kojima1998,BK1999}, the frequencies in TeVeS could be continuous spectrum limited to a certain range. The allowed frequencies are determined with the value of $\omega$ inside the star. Since the value of $\omega(r)$ is monotonically decreasing as $r$ is increasing, the minimum and maximum values of $\omega$ inside the star can be determined those values at the stellar surface and center, respectively, i.e., $\omega(\tilde{R})\le\omega\le\omega(0)$. Combing this evidence with Eq. (\ref{eq:sigma0}), the frequency of toroidal oscillation should be limited in the range of
\begin{equation}
  |\sigma_{\rm min}| \le |\sigma| \le |\sigma_{\rm max}|, \label{eq:range}
\end{equation}
where $\sigma_{\rm min}$ and $\sigma_{\rm max}$ are
\begin{align}
 \sigma_{\rm min} &= -m\tilde{\Omega} + \frac{2m}{\ell(\ell+1)}\left(\tilde{\Omega}-\omega(\tilde{R})\right), \\
 \sigma_{\rm max} &= -m\tilde{\Omega} + \frac{2m}{\ell(\ell+1)}\left(\tilde{\Omega}-\omega(0)\right).
\end{align}
From observational point of view, it is not sure whether this continuous spectrum could be observed or not. In fact, if one considers in GR the fast rotating star (or the coupling with the spheroidal oscillations), the frequencies of toroidal oscillations also show the discrete spectrum \cite{Erich2009,RSK2003}. So, the observed frequencies might be a single value in the range of Eq. (\ref{eq:range}). However, through the analysis as in this article, we believe that one can estimate how the gravitational theory affects on the frequency of toroidal oscillations.

%%%%%%%%%%%%%%%%%%%%%%%%%%%%%%%%%%%%%%%%%%%%%%%%%  FIGURE
%%%%%%%%%%%%%%%%%%%%%%%%%%%%%%%%%%%%%%%%%%%%%%%%% Figure 2
%\begin{figure}[htbp]
%\begin{center}
%\includegraphics[scale=0.45]{sigma-Omega-GR} 
%\end{center}
%\caption{%%
%Frequencies with $\ell=2$ in GR are plotted as a function of the stellar rotational frequency $\tilde{\Omega}$ for the stellar model with $M=1.4M_\odot$ and EOS A. With non-zero $m$, the solid lines denote the values of $\sigma_{\rm max}$, while the broken lines correspond to those of $\sigma_{\rm min}$.
%}%%
%\label{fig:sigma-Omega-GR}
%\end{figure}
%
%%%%%%%%%%%%%%%%%%%%%%%%%%%%%%%%%%%%%%%%%%%%%%%%%  FIGURE
%%%%%%%%%%%%%%%%%%%%%%%%%%%%%%%%%%%%%%%%%%%%%%%%% Figure 3
\begin{figure}[htbp]
\begin{center}
\includegraphics[scale=0.45]{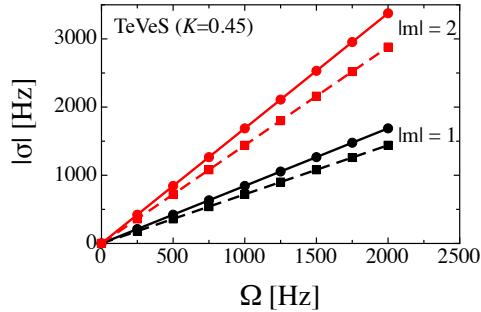} 
\end{center}
\caption{%%
Frequencies with $\ell=2$ in TeVeS with $K=0.45$ are plotted as a function of the stellar rotational frequency $\tilde{\Omega}$ for the stellar model with $M_{\rm ADM}=1.4M_\odot$ and EOS A, where the solid lines denote the values of $\sigma_{\rm max}$, while the broken lines correspond to those of $\sigma_{\rm min}$.
}%%
\label{fig:sigma-Omega-TeVeS}
\end{figure}
%

\begin{figure}[htbp]
\begin{center}
\begin{tabular}{cc}
\includegraphics[scale=0.45]{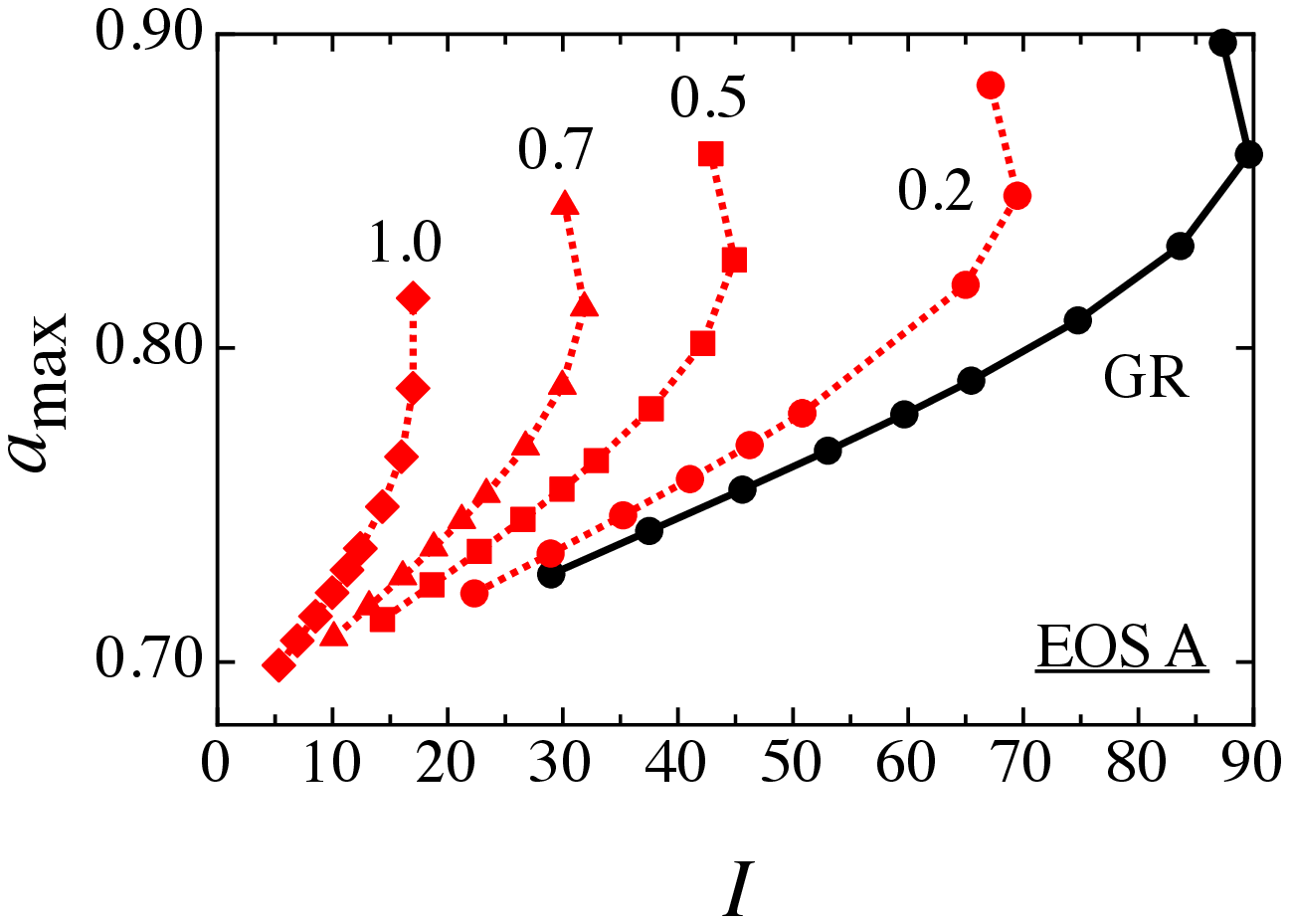} &
\includegraphics[scale=0.45]{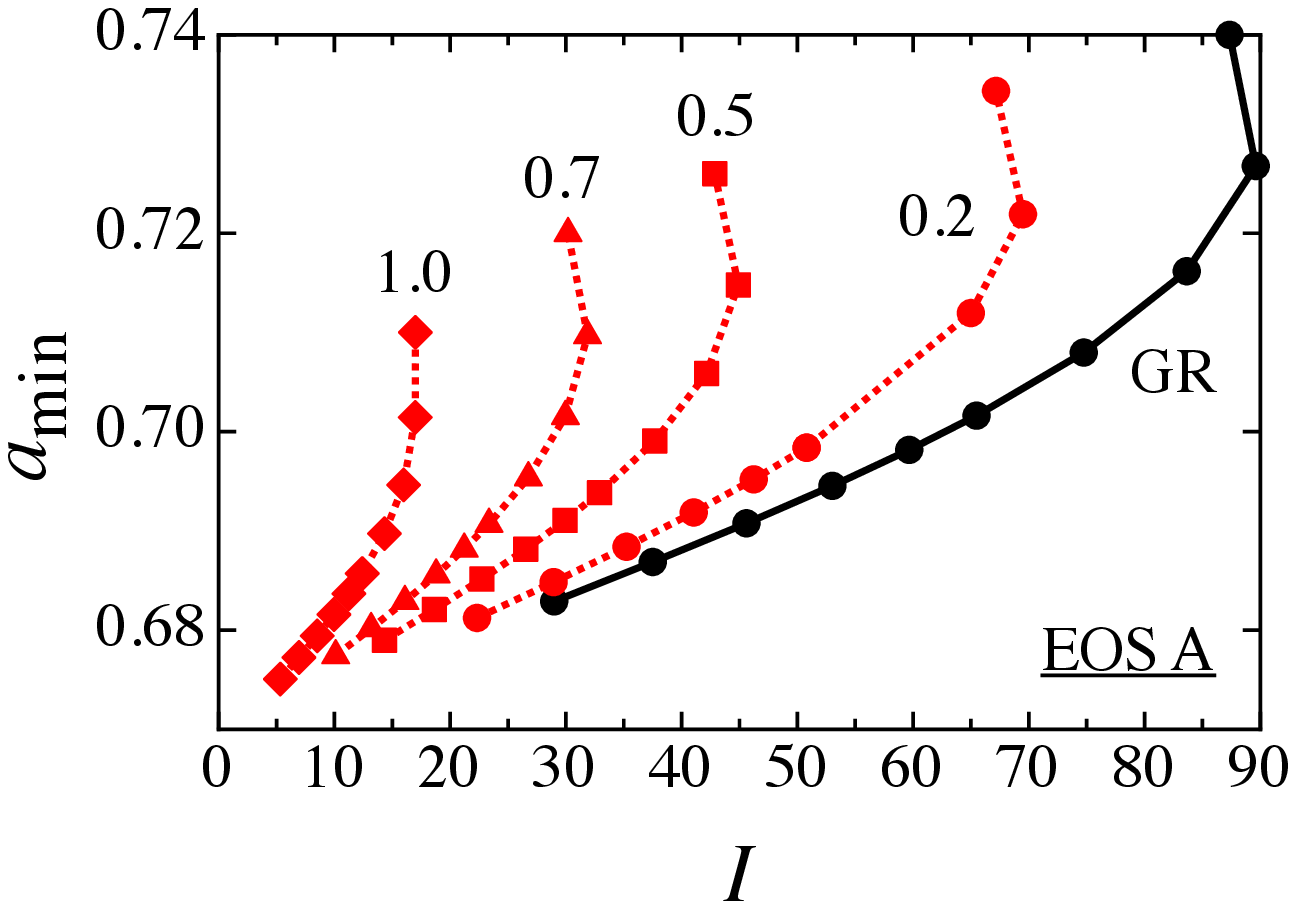} \\
\end{tabular}
\end{center}
\caption{%%
Values of $a_{\rm max}$ and $a_{\rm min}$ in GR and in TeVeS for EOS A as functions of $I$, where the solid lines correspond the results in GR and the dotted lines are those in TeVeS with different values of $K$, such as $K=0.2$, 0.5, 0.7, and 1.0.
}%%
\label{fig:aI_A}
\end{figure}

\begin{figure}[htbp]
\begin{center}
\begin{tabular}{cc}
\includegraphics[scale=0.45]{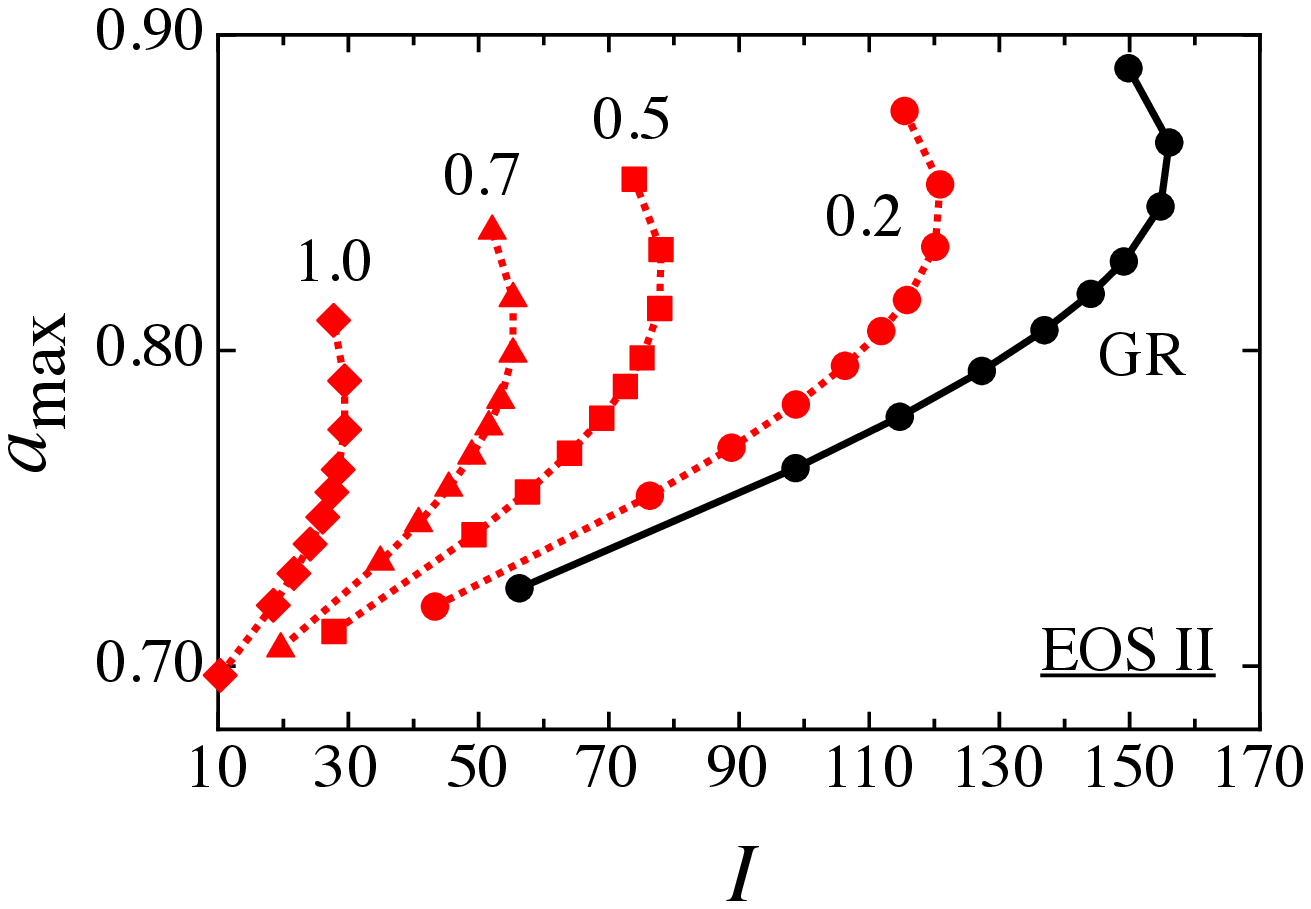} &
\includegraphics[scale=0.45]{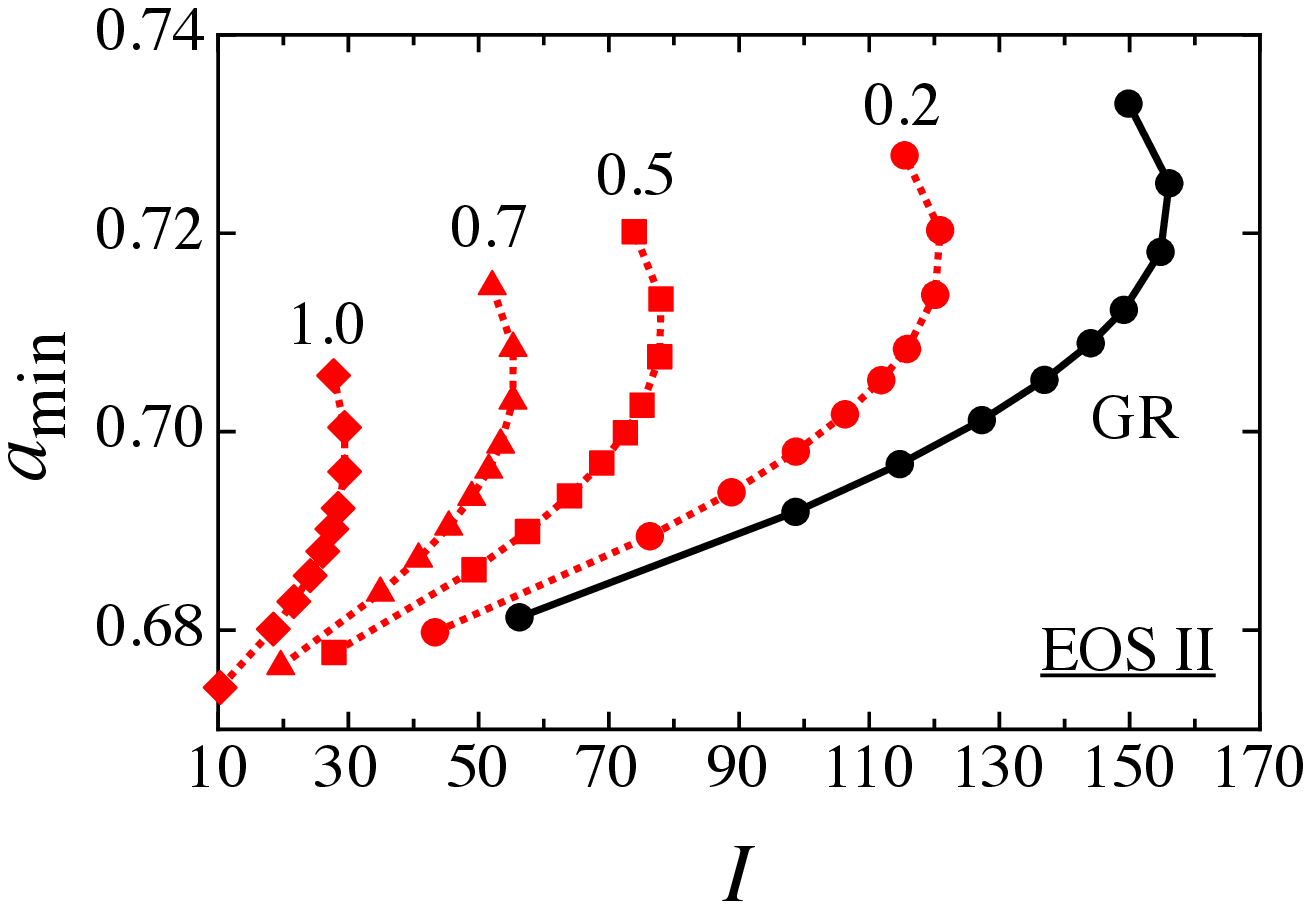} \\
\end{tabular}
\end{center}
\caption{%%
Similar to Fig. \ref{fig:aI_A}, but for EOS II.
}%%
\label{fig:aI_II}
\end{figure}

%For the comparison with the results in TeVeS, the frequencies of toroidal oscillations with $\ell=2$ in GR are shown as a function of angular velocity $\Omega$ in Fig. \ref{fig:sigma-Omega-GR}, where the adopted stellar model is constructed with EOS A and that mass is fixed to be $M=1.4M_\odot$. As shown this figure,
It is well-known that with the slowly rotating approximation, the frequencies of toroidal oscillations in GR are proportional to the angular velocity. On the other hand, the frequencies of toroidal oscillations in TeVeS with $\ell=2$ are shown in Fig. \ref{fig:sigma-Omega-TeVeS}, where the stellar mass for EOS A is fixed to be $M_{\rm ADM}=1.4M_\odot$ with $K=0.45$. From this figure, one can see that the frequencies even in TeVeS are proportional to $\tilde{\Omega}$ as same as in GR. Thus, we can define the new parameters, $a_{\rm max}$ and $a_{\rm min}$, as
\begin{equation}
 a_{\rm max} \equiv \left|\frac{\sigma_{\rm max}}{m\tilde{\Omega}}\right|
 \ \ \ {\rm and}\ \ \  
 a_{\rm min} \equiv \left|\frac{\sigma_{\rm min}}{m\tilde{\Omega}}\right|,
\end{equation}
which are independent of the value of $\tilde{\Omega}$ and $m$. Figs. \ref{fig:aI_A} and \ref{fig:aI_II} show these parameters in GR and in TeVeS as a function of $\tilde{I}$ defined by
\begin{equation}
 \tilde{I}\equiv \frac{\tilde{J}}{\tilde{\Omega}},
\end{equation}
where $\tilde{I}$ is constant and corresponding to the relativistic generalization of momentum of inertia for slowly rotating system \cite{Hartle1967,Sotani2010}. In these figures, as increasing the stellar mass $M_{\rm ADM}$ along each line, the values of $a_{\rm max}$ and $a_{\rm min}$ are also increasing. It should be noticed that the value of $\tilde{I}$ has not observed yet, but $\tilde{I}$ is an important parameter expressing the stellar configuration and that value might be determined if the phenomena of precession of rotating compact stars would be observed. From Figs. \ref{fig:aI_A} and \ref{fig:aI_II}, one can see that $a_{\rm max}$ and $a_{\rm min}$ are not so sensitive on the gravitational theory, while $\tilde{I}$ depends strongly on the gravitational theory. In practice, if the values of $a_{\rm max}$ and/or $a_{\rm min}$ would be determined, the value of $\tilde{I}$ could change around $80\%$ depending on the value of vector coupling $K$. Additionally, it could be found that the values of $a_{\rm max}$ and $a_{\rm min}$ are almost independent from the adopted equation of sate, but the corresponding value of $\tilde{I}$ depends strongly on the adopted equation of state. That is, via the detailed observations of $a_{\rm max}$ (or $a_{\min}$) and $\tilde{I}$, one could distinguish not only the gravitational theory in the strong-field regime but also the equation of state constructed the compact object.

%%%%%%%%%%%%%%%%%%%%%%%%%%%%%%%%%%%%%%%%%%%%%%%%%  FIGURE
%%%%%%%%%%%%%%%%%%%%%%%%%%%%%%%%%%%%%%%%%%%%%%%%% Figure 4ab
%

%%%%%%%%%%%%%%%%%%%%%%%%%%%%%%%%%%%%%%%%%%%%%%%%%  FIGURE
%%%%%%%%%%%%%%%%%%%%%%%%%%%%%%%%%%%%%%%%%%%%%%%%% Figure 5ab

\begin{figure}[htbp]
\begin{center}
\includegraphics[scale=0.45]{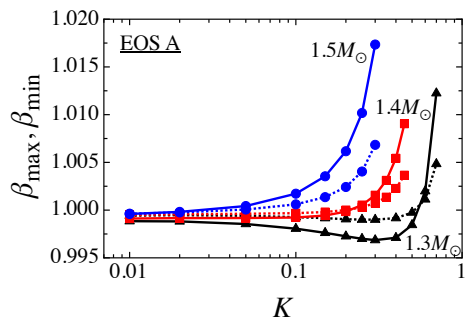} 
\end{center}
\caption{%%
Dependences of $\beta_{\rm max}$ and $\beta_{\rm min}$ on $K$ for EOS A, where the stellar masses are fixed to be $1.3M_\odot$ (traiangles), $1.4M_\odot$ (squares), and $1.5M_\odot$ (circles). The solid and dotted lines correspond to the values of $\beta_{\rm max}$ and $\beta_{\rm min}$, respectively.
}%%
\label{fig:beta_A}
\end{figure}

\begin{figure}[htbp]
\begin{center}
\includegraphics[scale=0.45]{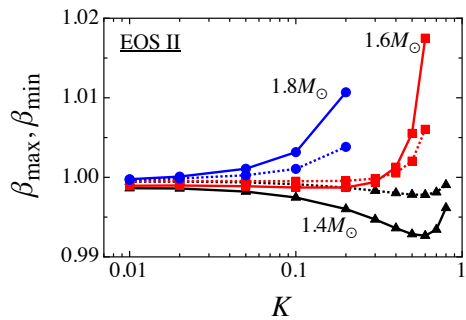} 
\end{center}
\caption{%%
Similar to Fig. \ref{fig:beta_A}, but for EOS II, where the stellar masses are fixed to be $1.4M_\odot$ (traiangles), $1.6M_\odot$ (squares), and $1.8M_\odot$ (circles).
}%%
\label{fig:beta_II}
\end{figure}

On the other hand, in Figs. \ref{fig:beta_A} and \ref{fig:beta_II}, the ratios of $a_{\rm max}$ and $a_{\rm min}$ in TeVeS to those in GR, which are defined by
\begin{equation}
 \beta_{\rm max} \equiv \frac{a_{\rm max}^{\rm (TeVeS)}}{a_{\rm max}^{\rm (GR)}}
 \ \ \ {\rm and} \ \ \ 
  \beta_{\rm min} \equiv \frac{a_{\rm min}^{\rm (TeVeS)}}{a_{\rm min}^{\rm (GR)}},
\end{equation}
are plotted as a function of the vector coupling $K$, where the stellar masses are fixed as $M_{\rm ADM}=1.3M_\odot$, $1.4M_\odot$, and $1.5M_\odot$ for EOS A, and $M_{\rm ADM}=1.4M_\odot$, $1.6M_\odot$, and $1.8M_\odot$ for EOS II. These figures show that with smaller $K$, the values of $\beta_{\rm max}$ and $\beta_{\rm min}$ become around 1 independently of stellar mass, while with larger $K$, one can see small deviation of $a_{\rm max}$ and $a_{\rm min}$ in TeVeS from those in GR at most a few \%. Furthermore, it is also found that as the stellar mass becomes larger, the values of $\beta_{\rm max}$ and $\beta_{\rm min}$ are larger. So, with massive stellar model and with larger value of $K$, it might be possible to distinguish the gravitational theory by observing the detailed toroidal oscillations with the help of the observation of stellar mass.

%%%%%%%%%%%%%%%%%%%%%%%%%%%%%%%%%%%%%%%%%%%%%%%%%  FIGURE
%%%%%%%%%%%%%%%%%%%%%%%%%%%%%%%%%%%%%%%%%%%%%%%%% Figure 6
%

%%%%%%%%%%%%%%%%%%%%%%%%%%%%%%%%%%%%%%%%%%%%%%%%%  FIGURE
%%%%%%%%%%%%%%%%%%%%%%%%%%%%%%%%%%%%%%%%%%%%%%%%% Figure 7
%

%%%%%%%%%%%%%%%%%%%%%%%%%%%%%%%%%%%%%%%%%%%%%%%%
\section{Conclusion}
\label{sec:IV}
%%%%%%%%%%%%%%%%%%%%%%%%%%%%%%%%%%%%%%%%%%%%%%%%

In order to examine the dependence of frequencies of toroidal oscillations on the gravitational theory, we consider such oscillations on the slowly rotating relativistic star in Tensor-Vector-Scalar (TeVeS) theory, where the angular velocity $\tilde{\Omega}$ is assumed constant. For this aim, we have derived the perturbation equations describing the toroidal oscillations of neutron stars in TeVeS and examine their specific frequencies as varying the vector coupling $K$ and stellar mass, where as a first step we focus only on the fluid oscillations and omit the perturbations of tensor, vector, and scalar fields (the Cowling approximation). We find that the perturbation equations in TeVeS can be written same form as in general relativity (GR). Therefore, similarly GR, the frequencies of toroidal oscillations show the continuous spectrum at least in the frame with the Cowling approximation. Comparing the results in GR and in TeVeS, one can see that although it seems that those frequencies are not so sensitive on the gravitational theory, the value of relativistic generalization of momentum of inertia depends strongly on the gravitational theory as well as the adopted equation of state. So, observing the frequencies of toroidal oscillations and momentum inertia with high accuracy might be able to reveal not only the gravitational theory in the strong-field regime but also the equation of state in the higher density region.

In this article, for simplicity, we adopt the Cowling approximation. That is, our consideration restricts only on the stellar oscillations. This means that we should do a more detailed study including the perturbations of other fields. With the oscillations of other fields, it could be possible to obtain the additional information about the different type of oscillation modes. That is, the observations of other oscillations can provide more accurate constraints on the gravitational theory in the strong-field regime. On the other hand, we should take into account the mode coupling between the toroidal and spheroidal oscillations, although such coupling is higher order effect. Considering this type of coupling, it is known in GR that the spectrum of toroidal oscillations could become discrete \cite{RSK2003}. The same can be expected in the case of TeVeS. If so, since the spheroidal oscillations depends strongly on the gravitational theory \cite{Sotani2009}, via the mode coupling, the frequencies of toroidal oscillations could also depend strongly on the gravitational theory. Thus, the observations of toroidal oscillations might become more important to distinguish the gravitational theory. Furthermore, it might be important to study the dependence of magnetic effects on the toroidal oscillations. In practice, the quasi-periodic oscillations have observed during the decaying tail of giant flares and these phenomena are believed to be related to the oscillations of strong magnetized neutron stars \cite{Sotani2007}. Taking into account the magnetic effects, one might be possible to obtain the further constraint in the gravitational theory.

%\newpage
%%%%%%%%%%%%%%%%%%%%%%%%%%%%%%%%%%%%%%%%%%%%%%%%
\acknowledgments
%%%%%%%%%%%%%%%%%%%%%%%%%%%%%%%%%%%%%%%%%%%%%%%%
We are grateful to Kostas D. Kokkotas and Miltos Vavoulidis for their warm hospitality and fruitful discussions, and also to the referees for careful reading my manuscript and giving a valuable comments.
%This work was partially supported by   .

%\appendix
%%%%%%%%%%%%%%%%%%%%%%%%%%%%%%%%%%%%%%%%%%%%%%%%
%\section{}   % Appendix A
%\label{sec:appendix_1}
%%%%%%%%%%%%%%%%%%%%%%%%%%%%%%%%%%%%%%%%%%%%%%%%

%%%%%%%%%%%%%%%%%%%%%%%%%%%%%%%%%%%%%%%%%%%%%%%%

\end{document}